\documentstyle[12pt]{article}

\begin{document}

\title{{\LARGE {\bf On Modifications of the Sp(2) Covariant Superfield Quantization}%
}}
\date{}
\author{{\sc D.M. Gitman\thanks{%
e-mail: gitman@dfn.if.usp.br} and P.Yu. Moshin\thanks{{Tomsk State
Pedagogical University, 634041 Tomsk, Russia; }e-mail: moshin@dfn.if.usp.br}}
\\
\\
{Instituto de F{\'{\i}}sica, Universidade de S\~{a}o Paulo,}\\
{Caixa Postal 66318-CEP, 05315-970 S\~{a}o Paulo, S.P., Brazil}}
\maketitle

\begin{abstract}
{\small We propose a modification of the $Sp$(2) covariant superfield
quantization to realize a superalgebra of generating operators isomorphic to
the massless limit of the corresponding superalgebra of the $osp$(1,2)
covariant formalism. The modified scheme ensures the compatibility of the
superalgebra of generating operators with extended BRST symmetry without
imposing restrictions eliminating superfield components from the quantum
action. The formalism coincides with the $Sp$(2) covariant superfield scheme
and with the massless limit of the $osp$(1,2) covariant quantization in
particular cases of gauge-fixing and solutions of the quantum master
equations. }
\end{abstract}

\section{Introduction}

The covariant quantization of gauge theories is based on the concept of
quantum master equations, realized in terms of the corresponding generating
operators and antibrackets (see, e.g., \cite{blt,3pl,glm}). Quantum master
equations encode the presence of BRST\ symmetry, being a global
supersymmetry of the integrand in the vacuum functional. This symmetry was
discovered in Yang--Mills theories and then generalized to extended BRST
symmetry, which combines BRST \cite{brst} and antiBRST \cite{antibrst}
transformations. Extended BRST symmetry permitted to find a superspace
description \cite{bpt} of quantum Yang--Mills theories, where this symmetry
was realized in terms of supertranslations along additional anticommuting
coordinates.

In general gauge theories (with an arbitrary gauge algebra and stage of
reducibility), extended BRST\ symmetry is realized within the $Sp(2)$
covariant quantization scheme \cite{blt} and its different modifications,
e.g., \cite{3pl,glm}, including the $osp(1,2)$ covariant formalism \cite{glm}%
. The quantization scheme \cite{blt} describes the structure of the complete
configuration space of a theory in terms of irreducible representations of
the group $Sp(2)${\it .} The scheme \cite{glm} modifies the formalism \cite
{blt} in a way which ensures the $Sp(2)$ invariance of a theory by imposing
on the quantum action a set of master equations and analogous subsidiary
conditions with the corresponding generating operators subject to a
superalgebra isomorphic to $osp(1,2)$. The superalgebra of the generating
operators \cite{glm} depends on a mass parameter, inherited by the quantum
action. The incorporated mass is intended to serve as a regularization
parameter of an $Sp(2)$\ invariant renormalization of the quantum theory 
\cite{glm}. The superfield versions \cite{l,gm,lm2} of the $Sp(2)$ and $%
osp(1,2)$ covariant schemes realize superspace formulations of extended
BRST\ symmetry in general gauge theories.

The superfield formalism \cite{l} combines the variables used in the $Sp(2)$
covariant scheme \cite{blt} into a set of superfields and supersources
defined in a superspace with two anticommuting coordinates. The quantum
action is given by a functional of superfields and supersources, which makes
it possible to present extended BRST symmetry in terms of supertranslations
and transformations generated by superfield antibrackets \cite{l}{\it .}{\em %
\ }A possible arbitrariness in the form of superfield antibrackets
compatible with the superspace interpretation of extended BRST\ symmetry was
also examined in \cite{l}.

There are two alternative superfield formulations \cite{gm,lm2} of the $%
osp(1,2)$ covariant scheme \cite{glm}. Both formulations, constructed along
the lines of \cite{l}, are not free from difficulties. Thus, in \cite{gm}
there exists an inconsistency (see \cite{lm2} for a detailed discussion)
between the form of superfield antibrackets and the extended BRST symmetry
realized in terms of supertranslations. In \cite{lm2}, this problem is
solved at the cost of eliminating some superfield components from the
quantum action, which implies that the extended BRST\ symmetry in \cite{lm2}
is not entirely controlled by the quantum master equations.

In \cite{lm2}, it was also remarked that a consistent superspace formulation
of the $osp(1,2)$\ covariant approach should contain the $Sp(2)$\ covariant
superfield scheme \cite{l} in the massless limit. On the one hand, this can
be explained by the fact that the massless limit of the $osp(1,2)$ covariant
scheme contains the original $Sp(2)$ covariant formalism, regarded in a
special case of gauge-fixing and solutions of the master equations. On the
other hand, the superfield description of the $Sp(2)$\ covariant scheme \cite
{l} realizes the only form of superfield antibrackets which respects the
superspace interpretation of extended BRST\ symmetry without additional
restrictions on the quantum action. A non-trivial problem facing the
proposal \cite{lm2} is to ensure a compatibility of the superfield
antibrackets \cite{l} with the $osp(1,2)$ superalgebra of generating
operators (see also \cite{gm,lm2}).

To advance in the solution of this problem, we demonstrate the existence of
a superfield scheme which can be identified with a massless limit suggested
in \cite{lm2}. To this end, we propose a superfield scheme based on a set of
generating operators which form a superalgebra isomorphic to the massless
limit of the superalgebra realized in the $osp(1,2)$ covariant scheme \cite
{glm}. The choice of generating operators is consistent with the superfield
antibrackets \cite{l} and the superspace form of extended BRST\ symmetry,
without imposing restrictions eliminating superfield components.{\em \ }The
formalism contains the $Sp(2)$ covariant superfield scheme \cite{l} and the
massless limit of the $osp(1,2)$ covariant approach \cite{glm}. Massive
extensions of the proposed formalism may provide a key to constructing a
superfield $osp(1,2)$ covariant scheme free from the problems that remain in 
\cite{gm,lm2}.

The paper is organized as follows. In Section 2, we introduce the main
definitions. In Section 3, we formulate the quantization rules. In Section
4, we discuss the relation of the proposed formalism to the quantization
schemes \cite{glm,l}. In Section 5, we summarize the results and make
concluding remarks.

We use the notation adopted in \cite{glm,l}. Derivatives with respect to
(super)sources and antifields are taken from the left, and those with
respect to (super)fields, from the right. Left derivatives with respect to
(super)fields are labeled by the subscript {\it ``l''}. Integration over
superfields and supersources is understood as integration over their
components.

\section{Main Definitions}

Let us consider a superspace $(x^{\mu },\theta ^{a})$, where $x^{\mu }$ are
space-time coordinates, and $\theta ^{a}$ is an $Sp(2)$ doublet of
anticommuting coordinates. Notice that any function $f(\theta )$ has a
component representation, 
\[
f(\theta )=f_{0}+\theta ^{a}f_{a}+\theta ^{2}f_{3},\;\;\;\theta ^{2}\equiv 
\frac{1}{2}\theta _{a}\theta ^{a}, 
\]
and an integral representation, 
\[
f(\theta )=\int d^{2}\theta ^{\prime }\,\delta (\theta ^{\prime }-\theta
)f(\theta ^{\prime }),\;\;\;\delta (\theta ^{\prime }-\theta )=(\theta
^{\prime }-\theta )^{2}, 
\]
where raising and lowering the $Sp(2)$ indices is performed by the rule $%
\theta ^{a}=\varepsilon ^{ab}\theta _{b}$, $\theta _{a}=\varepsilon
_{ab}\theta ^{b}$, with $\varepsilon ^{ab}$ being a constant antisymmetric
tensor, $\varepsilon ^{12}=1$, and integration over $\theta ^{a}$ is given
by 
\[
\int d^{2}\theta =0,\;\;\int d^{2}\theta \;\theta ^{a}=0,\;\;\int
d^{2}\theta \;\theta ^{a}\theta ^{b}=\varepsilon ^{ab}. 
\]
In particular, for any function $f(\theta )$ we have 
\[
\int d^{2}\theta \;\frac{\partial f(\theta )}{\partial \theta ^{a}}=0, 
\]
which implies the property of integration by parts 
\begin{equation}
\int d^{2}\theta \;\frac{\partial f(\theta )}{\partial \theta ^{a}}g(\theta
)=-\!\int \!d^{2}\theta (-1)^{\varepsilon (f)}f(\theta )\frac{\partial
g(\theta )}{\partial \theta ^{a}},  \label{byparts}
\end{equation}
where derivatives with respect to $\theta ^{a}$ are taken from the left.

We now introduce a set of superfields $\Phi ^{A}{(\theta )}$, $\varepsilon
(\Phi ^{A})=\varepsilon _{A}$, with the boundary condition 
\[
\left. \Phi ^{A}(\theta )\right| _{\theta =0}=\phi ^{A}, 
\]
and a set of supersources $\bar{\Phi}_{A}{(\theta )}$ of the same Grassmann
parity, $\varepsilon (\bar{\Phi}_{A})=\varepsilon _{A}$. The structure \cite
{blt} of the complete configuration space $\phi ^{A}$ of a general gauge
theory of $L$-stage reducibility is given by 
\begin{equation}
\phi ^{A}=(A^{i},B^{\alpha _{s}|a_{1}\cdots a_{s}},C^{\alpha
_{s}|a_{0}\cdots a_{s}}),\;\;s=0,\ldots ,L,  \label{ind}
\end{equation}
where $A^{i}$ are the initial classical fields, while $B^{\alpha
_{s}|a_{1}\cdots a_{s}}$, $C^{\alpha _{s}|a_{0}\cdots a_{s}}$ are the
pyramids of auxiliary and (anti)ghost fields, being completely symmetric $%
Sp(2)$ tensors of rank $s$ and $s+1$, respectively.

For arbitrary functionals $F=F(\Phi ,\bar{\Phi})$, $G=G(\Phi ,\bar{\Phi})$,
we define the superbracket operations $(\,\,,\,)^{a}$ and $%
\{\,\,,\,\}_{\alpha }$ 
\begin{eqnarray}
\!\!\!\!\!\!\!\!\!\!\!\! &&(F,G)^{a}=\int d^{2}\theta \left\{ \frac{\delta F%
}{\delta \Phi ^{A}(\theta )}\frac{\partial }{\partial \theta _{a}}\frac{%
\delta G}{\delta \bar{\Phi}_{A}(\theta )}(-1)^{\varepsilon
_{A}+1}-(-1)^{(\varepsilon (F)+1)(\varepsilon (G)+1)}(F\leftrightarrow
G)\right\} ,  \nonumber \\
\!\!\!\!\!\!\!\!\!\!\!\! &&\{F,G\}_{\alpha }=-\int d^{2}\theta \Bigg\{%
(\sigma _{\alpha })_{B}^{~~\!A}\left[ \frac{\partial ^{2}}{\partial \theta
^{2}}\left( \frac{\delta F}{\delta \Phi ^{A}(\theta )}\right) \theta ^{2}-%
\frac{\partial ^{2}}{\partial \theta ^{2}}\left( \frac{\delta F}{\delta \Phi
^{A}(\theta )}\theta ^{2}\right) \right] \frac{\delta G}{\delta \bar{\Phi}%
_{B}(\theta )}  \nonumber \\
\!\!\!\!\!\!\!\!\!\!\!\! &&+\,\frac{\partial ^{2}}{\partial \theta ^{2}}%
\left( \frac{\delta F}{\delta \Phi ^{A}(\theta )}\,\theta _{b}(\sigma
_{\alpha })_{\;\;a}^{b}+\frac{\delta F}{\delta \Phi ^{B}(\theta )}\,\theta
_{a}(\sigma _{\alpha })_{\;\;A}^{B}\right) \theta ^{a}\frac{\delta G}{\delta 
\bar{\Phi}_{A}(\theta )}+(-1)^{\varepsilon (F)\varepsilon
(G)}(F\leftrightarrow G)\Bigg\},  \label{bracket}
\end{eqnarray}
where 
\[
\frac{\partial ^{2}}{\partial \theta ^{2}}\equiv \frac{1}{2}\varepsilon ^{ab}%
\frac{\partial }{\partial \theta ^{b}}\frac{\partial }{\partial \theta ^{a}}%
\;. 
\]
Notice the properties of derivatives 
\[
\frac{\delta _{l}\Phi ^{A}(\theta )}{\delta \Phi ^{B}(\theta ^{^{\prime }})}=%
\frac{\delta \Phi ^{A}(\theta )}{\delta \Phi ^{B}(\theta ^{^{\prime }})}%
=\delta (\theta ^{^{\prime }}-\theta )\delta _{B}^{A},\;\;\;\;\frac{\delta 
\bar{\Phi}_{A}(\theta )}{\delta \bar{\Phi}_{B}(\theta ^{^{\prime }})}=\delta
(\theta ^{^{\prime }}-\theta )\delta _{A}^{B}. 
\]
In (\ref{bracket}), the matrices $(\sigma _{\alpha })_{A}^{~~\!B}\equiv
-(\sigma _{\alpha })_{~~\!A}^{B}$, with the indices (\ref{ind}), are given
by 
\begin{equation}
(\sigma _{\alpha })_{~~\!A}^{B}=(\sigma _{\alpha })_{~a}^{b}(P_{\pm
})_{Ab}^{Ba}.  \label{sigma}
\end{equation}
Here, $(\sigma _{\alpha })_{~a}^{b}$, with $\alpha =(0,+,-)$, stands for a
set of matrices which possess the properties 
\[
(\sigma _{\alpha })_{~a}^{b}=-(\sigma _{\alpha })_{a}^{~b},\;\;\;(\sigma
_{\alpha })^{ab}=\varepsilon ^{ac}(\sigma _{\alpha })_{c}^{~b}=(\sigma
_{\alpha })_{~c}^{a}\,\varepsilon ^{cb}=\varepsilon ^{ac}(\sigma _{\alpha
})_{cd}\,\varepsilon ^{db},\;\;\;(\sigma _{\alpha })^{ab}=(\sigma _{\alpha
})^{ba}, 
\]
\begin{equation}
(\sigma _{\alpha })_{a}^{~a}=(\sigma _{\alpha
})_{~a}^{a}=0,\;\;\;\varepsilon ^{ad}\delta _{c}^{b}+\varepsilon ^{bd}\delta
_{c}^{a}=-(\sigma ^{\alpha })^{ab}(\sigma _{\alpha })_{~c}^{d}
\label{propsigma}
\end{equation}
and form the algebra $sl(2)$ 
\[
\sigma _{\alpha }\sigma _{\beta }=g_{\alpha \beta }+\hbox{$\frac{1}{2}$}%
\epsilon _{\alpha \beta \gamma }\sigma ^{\gamma },\;\;\;\sigma ^{\alpha
}=g^{\alpha \beta }\sigma _{\beta },\;\;\;\hbox{Tr}(\sigma _{\alpha }\sigma
_{\beta })=2g_{\alpha \beta }, 
\]
\[
g^{\alpha \beta }=\left( 
\begin{array}{ccc}
1 & 0 & 0 \\ 
0 & 0 & 2 \\ 
0 & 2 & 0
\end{array}
\right) ,\;\;\;g^{\alpha \gamma }g_{\gamma \beta }=\delta _{\beta }^{\alpha
}, 
\]
with $\epsilon _{\alpha \beta \gamma }$ being an antisymmetric tensor, $%
\epsilon _{0+-}=1$.

In (\ref{sigma}), the matrices $(P_{\pm })_{Ab}^{Ba}$ are given by 
\[
(P_{\mp })_{Ab}^{Ba}=(P_{\pm })_{Ab}^{Ba}-(P_{\pm })_{A}^{B}\delta
_{b}^{a}+\delta _{A}^{B}\delta _{b}^{a},\;\;\;(P_{\pm })_{A}^{B}=\delta
_{a}^{b}(P_{\pm })_{Ab}^{Ba}, 
\]
where 
\[
(P_{+})_{Ab}^{Ba}=\left\{ 
\begin{array}{ll}
\delta _{j}^{i}\delta _{b}^{a} & A=i,B=j, \\ 
\delta _{\alpha _{s}}^{\beta _{s}}(s+1)S_{a_{1}\cdots a_{s}b}^{b_{1}\cdots
b_{s}a} & A=\alpha _{s}|a_{1}\cdots a_{s},\;B=\beta _{s}|b_{1}\cdots b_{s},
\\ 
\delta _{\alpha _{s}}^{\beta _{s}}(s+2)S_{a_{0}\cdots a_{s}b}^{b_{0}\cdots
b_{s}a} & A=\alpha _{s}|a_{0}\cdots a_{s},\;B=\beta _{s}|b_{0}\cdots b_{s},
\\ 
0 & \mbox{otherwise}.
\end{array}
\right. 
\]
Here, $S_{a_{0}\cdots a_{s}b}^{b_{0}\cdots b_{s}a}$ is a symmetrizer ($X^{a}$
being independent bosonic variables) 
\[
S_{a_{0}\cdots a_{s}b}^{b_{0}\cdots b_{s}a}\equiv \frac{1}{(s+2)!}\frac{%
\partial }{\partial X^{a_{0}}}\cdots \frac{\partial }{\partial X^{a_{s}}}%
\frac{\partial }{\partial X^{b}}X^{a}X^{b_{s}}\cdots X^{b_{0}}, 
\]
with the properties 
\begin{eqnarray*}
S_{a_{0}\cdots a_{s}b}^{b_{0}\cdots b_{s}a} &=&\frac{1}{s+2}\left(
\sum_{r=0}^{s}\delta _{a_{0}}^{b_{r}}S_{a_{1}\cdots a_{s}b}^{b_{0}\cdots
b_{r-1}b_{r+1}\cdots b_{s}a}+\frac{1}{s+1}\sum_{r=0}^{s}\delta
_{a_{0}}^{a}\delta _{b}^{b_{r}}S_{a_{1}\cdots a_{s}}^{b_{0}\cdots
b_{r-1}b_{r+1}\cdots b_{s}}\right) , \\
S_{a_{0}\cdots a_{s}}^{b_{0}\cdots b_{s}} &=&\frac{1}{s+1}%
\sum_{r=0}^{s}\delta _{a_{0}}^{b_{r}}S_{a_{1}\cdots a_{s}}^{b_{0}\cdots
b_{r-1}b_{r+1}\cdots b_{s}}.
\end{eqnarray*}
From the above definitions follow the properties \cite{glm} 
\[
(P_{\mp })_{Cd}^{Ab}(P_{\pm })_{Ba}^{Cd}=0,\;\;\;\;\varepsilon ^{ad}(P_{\pm
})_{Ad}^{Bb}+\varepsilon ^{bd}(P_{\pm })_{Ad}^{Ba}=-(\sigma ^{\alpha
})^{ab}(\sigma _{\alpha })_{~A}^{B}, 
\]
\[
\varepsilon ^{ad}(P_{\pm })_{Ac}^{Bb}+\varepsilon ^{bd}(P_{\pm
})_{Ac}^{Ba}-(\sigma ^{\alpha })^{ab}(\sigma _{\alpha })_{~c}^{e}(P_{\mp
})_{Ae}^{Bd}=-(\sigma ^{\alpha })^{ab}\bigr(
(\sigma _{\alpha })_{~c}^{d}\delta _{A}^{B}+\delta _{c}^{d}(\sigma _{\alpha
})_{~A}^{B}\bigr)
. 
\]

We now introduce a set of first-order operators $V^{a}$, $U^{a}$ (odd) and $%
V_{\alpha }$, $U_{\alpha }$ (even), 
\begin{eqnarray}
V^{a} &=&\int d^{2}\theta \frac{\partial \bar{\Phi}_{A}(\theta )}{\partial
\theta _{a}}\frac{\delta }{\delta \bar{\Phi}_{A}(\theta )},  \nonumber \\
U^{a} &=&\int d^{2}\theta \frac{\partial {\Phi }^{A}(\theta )}{\partial
\theta _{a}}\frac{\delta _{l}}{\delta {\Phi }^{A}(\theta )},  \nonumber \\
V_{\alpha } &=&\int d^{2}\theta \left( \bar{\Phi}_{B}(\sigma _{\alpha
})_{~~\!A}^{B}\frac{\delta }{\delta \bar{\Phi}_{A}(\theta )}-\frac{\partial
^{2}}{\partial \theta ^{2}}\left( \bar{\Phi}_{A}(\theta )\theta _{b}\right)
(\sigma _{\alpha })_{~a}^{b}\theta ^{a}\frac{\delta }{\delta \bar{\Phi}%
_{A}(\theta )}\right) ,  \nonumber \\
U_{\alpha } &=&\int d^{2}\theta \left( \Phi ^{A}(\sigma _{\alpha
})_{A}^{~~\!B}\frac{\delta _{l}}{\delta \Phi ^{B}(\theta )}+\frac{\partial
^{2}}{\partial \theta ^{2}}\left( \Phi ^{A}(\theta )\theta ^{a}\right)
(\sigma _{\alpha })_{a}^{~b}\theta _{b}\frac{\delta _{l}}{\delta \Phi
^{A}(\theta )}\right) .  \label{U&V}
\end{eqnarray}
These operators obey a superalgebra with the following non-trivial
(anti)com\-mu\-ta\-tion relations: 
\[
\lbrack V_{\alpha },V_{\beta }]=\epsilon _{\alpha \beta }^{~~~\!\gamma
}V_{\gamma },\;\;\;[V_{\alpha },V^{a}]=V^{b}(\sigma _{\alpha
})_{b}^{~a},\;\;\;\{V^{a},V^{b}\}=0, 
\]

\begin{equation}
\lbrack U_{\alpha },U_{\beta }]=-\epsilon _{\alpha \beta }^{~~~\!\gamma
}U_{\gamma },\;\;\;[U_{\alpha },U^{a}]=-U^{b}(\sigma _{\alpha
})_{b}^{~a},\;\;\;\{U^{a},U^{b}\}=0.  \label{uvalg}
\end{equation}

We also introduce a set of second-order operators $\Delta ^{a}$ (odd) and $%
\Delta _{\alpha }$ (even) 
\begin{eqnarray}
\Delta ^{a} &=&-\int d^{2}\theta \frac{\delta _{l}}{\delta \Phi ^{A}(\theta )%
}\frac{\partial }{\partial \theta _{a}}\frac{\delta }{\delta \bar{\Phi}%
_{A}(\theta )},  \nonumber \\
\Delta _{\alpha } &=&(-1)^{\varepsilon _{A}+1}\int d^{2}\theta \,\Bigg\{%
(\sigma _{\alpha })_{B}^{~~\!A}\left[ \frac{\partial ^{2}}{\partial \theta
^{2}}\left( \frac{\delta _{l}}{\delta \Phi ^{A}(\theta )}\right) \theta ^{2}-%
\frac{\partial ^{2}}{\partial \theta ^{2}}\left( \frac{\delta _{l}}{\delta
\Phi ^{A}(\theta )}\theta ^{2}\right) \right]  \nonumber \\
&&+\frac{\partial ^{2}}{\partial \theta ^{2}}\left( \frac{\delta _{l}}{%
\delta \Phi ^{B}(\theta )}\,\theta _{b}(\sigma _{\alpha })_{\;\;a}^{b}+\frac{%
\delta _{l}}{\delta \Phi ^{A}(\theta )}\,\theta _{a}(\sigma _{\alpha
})_{\;\;B}^{A}\right) \theta ^{a}\Bigg\}\frac{\delta }{\delta \bar{\Phi}%
_{B}(\theta )}\,.  \label{DeltaaExSQ2}
\end{eqnarray}
These operators possess the algebraic properties 
\begin{equation}
\lbrack \Delta _{\alpha },\Delta _{\beta }]=0,\;\;\;\{\Delta ^{a},\Delta
^{b}\}=0,\;\;\;[\Delta _{\alpha },\Delta ^{a}]=0,  \label{deltalg}
\end{equation}
\begin{eqnarray}
{\lbrack }\Delta _{\alpha },V_{\beta }{]}+{[}V_{\alpha },\Delta _{\beta }{]}
&=&\epsilon _{\alpha \beta }^{~~~\!\gamma }\Delta _{\gamma },  \nonumber \\
\{\Delta ^{a},V^{b}\}+\{V^{a},\Delta ^{b}\} &=&0,  \nonumber \\
{\lbrack }\Delta _{\alpha },V^{a}{]}+{[}V_{\alpha },\Delta ^{a}{]} &=&\Delta
^{b}(\sigma _{\alpha })_{b}^{~a}.  \label{deltalg2}
\end{eqnarray}
From (\ref{DeltaaExSQ2}) it follows that the action of the operators $\Delta
^{a}$ and $\Delta _{\alpha }$ on the product of two functionals defines the
superbracket operations (\ref{bracket}), namely, 
\begin{eqnarray}
\Delta _{\alpha }(FG) &=&(\Delta _{\alpha }F)G+F(\Delta _{\alpha
}G)+\{F,G\}_{\alpha },  \nonumber \\
\Delta ^{a}(FG) &=&(\Delta ^{a}F)G+F(\Delta ^{a}G)(-1)^{\varepsilon
(F)}+(F,G)^{a}(-1)^{\varepsilon (F)}.  \label{deltabracket}
\end{eqnarray}
Using the relations (\ref{deltalg}), (\ref{deltalg2}), (\ref{deltabracket}),
one can establish the properties of the superbrackets (\ref{bracket}) at the
algebraic level \cite{glm}.

Finally, we introduce the operators 
\[
\bar{\Delta}^{a}\equiv \Delta ^{a}+\frac{i}{\hbar }V^{a},\;\;\;\bar{\Delta}%
_{\alpha }\equiv \Delta _{\alpha }+\frac{i}{\hbar }V_{\alpha }. 
\]
From (\ref{uvalg}), (\ref{deltalg}), (\ref{deltalg2}) it follows that these
operators obey the superalgebra 
\begin{eqnarray*}
{\lbrack }\bar{\Delta}_{\alpha },\bar{\Delta}_{\beta }{]} &=&(i/\hbar
)\epsilon _{\alpha \beta }^{~~~\!\gamma }\bar{\Delta}_{\gamma }, \\
{\lbrack }\bar{\Delta}_{\alpha },\bar{\Delta}^{a}{]} &=&(i/\hbar )\bar{\Delta%
}^{b}(\sigma _{\alpha })_{b}^{~a}, \\
\{\bar{\Delta}^{a},\bar{\Delta}^{b}\} &=&0,
\end{eqnarray*}
isomorphic to the massless limit of the superalgebra of generating operators
used in the $osp(1,2)$-covariant quantization scheme \cite{glm}.

\section{Quantization Rules}

Let us define the vacuum functional $Z$ as the following path integral: 
\begin{equation}
Z=\int d\Phi \,d\bar{\Phi}\,\exp \left[ \frac{i}{\hbar }\left( W(\Phi ,\bar{%
\Phi})-\frac{1}{2}\varepsilon _{ab}U^{a}U^{b}F(\Phi )+\bar{\Phi}\Phi \right) %
\right] .  \label{ZExSQ}
\end{equation}
Here, $W=W(\Phi ,\bar{\Phi})$ is the quantum action, satisfying the boundary
condition 
\[
\left. W\right| _{\bar{\Phi}=\hbar =0}={\cal S}, 
\]
where ${\cal S}={\cal S}(A)$ is the action of the original gauge theory. The
quantum action $W$ is subject to the master equations 
\begin{equation}
\bar{\Delta}^{a}\exp \left( \frac{i}{\hbar }W\right) =0,  \label{GEqExSQ}
\end{equation}
and the subsidiary conditions 
\begin{equation}
\bar{\Delta}_{\alpha }\exp \left( \frac{i}{\hbar }W\right) =0,
\label{GEqExSQ2}
\end{equation}
with $\bar{\Delta}^{a}$ and $\bar{\Delta}_{\alpha }$ given by (\ref
{DeltaaExSQ2}). Equations (\ref{GEqExSQ}) and (\ref{GEqExSQ2}) are
equivalent to 
\begin{eqnarray}
\frac{1}{2}(W,W)^{a}+V^{a}W &=&i\hbar \Delta ^{a}W,  \label{GEq1ExSQ} \\
\frac{1}{2}\{W,W\}_{\alpha }+V_{\alpha }W &=&i\hbar \Delta _{\alpha }W,
\label{GEq1ExSQ2}
\end{eqnarray}
where the superbrackets $(\,\,,\,)^{a}$, $\{\,\,,\,\}_{\alpha }$ and the
operators $V^{a}$, $V_{\alpha }$, $\Delta ^{a}$, $\Delta _{\alpha }$ are
defined by (\ref{bracket}), (\ref{U&V}), (\ref{DeltaaExSQ2}). The quantum
action $W$ is also assumed to be an admissible solution of (\ref{GEq1ExSQ})
and (\ref{GEq1ExSQ2}). Namely, it is subject to the restriction 
\begin{equation}
\int d^{2}\theta \,\theta ^{2}\left( \frac{\delta W}{\delta \bar{\Phi}%
_{A}(\theta )}+\Phi ^{A}(\theta )\right) =0.  \label{IV51}
\end{equation}
In (\ref{ZExSQ}), $\bar{\Phi}\Phi $ is a functional of the form 
\begin{equation}
\bar{\Phi}\Phi =\int d^{2}\theta \,\bar{\Phi}_{A}(\theta )\Phi ^{A}(\theta ),
\label{phi_phi}
\end{equation}
while $F(\Phi )$ is a gauge-fixing Boson restricted by the conditions 
\begin{equation}
U_{\alpha }F(\Phi )=0,  \label{cond2}
\end{equation}
where $U_{\alpha }$ are the operators (\ref{U&V}).

An important property of the integrand in (\ref{ZExSQ}) is its invariance
under the following transformations: 
\begin{eqnarray}
\delta \Phi ^{A}(\theta ) &=&\mu _{a}U^{a}\Phi ^{A}(\theta ),\;\;\;\delta 
\bar{\Phi}_{A}(\theta )=\mu _{a}V^{a}\bar{\Phi}_{A}(\theta )+\mu _{a}(W,\bar{%
\Phi}_{A}(\theta ))^{a},  \label{BRSTExSQ} \\
\delta \Phi ^{A}(\theta ) &=&\mu ^{\alpha }U_{\alpha }\Phi ^{A}(\theta
),\;\;\;\delta \bar{\Phi}_{A}(\theta )=\mu ^{\alpha }V_{\alpha }\bar{\Phi}%
_{A}(\theta )+\mu ^{\alpha }\{W,\bar{\Phi}_{A}(\theta )\}_{\alpha },
\label{BRSTExSQ2}
\end{eqnarray}
where $U^{a}$ are operators given by (\ref{U&V}), while $\mu _{a}$ and $\mu
^{\alpha }$ are constant (anti)commuting parameters, $\varepsilon (\mu
_{a})=1$, $\varepsilon (\mu ^{\alpha })=0$. The validity of the symmetry
transformations (\ref{BRSTExSQ}), (\ref{BRSTExSQ2}) follows from the master
equations (\ref{GEq1ExSQ}), (\ref{GEq1ExSQ2}) and the conditions (\ref{cond2}%
) for the gauge-fixing Boson, with allowance for integration by parts (\ref
{byparts}) and the algebraic properties (\ref{uvalg}).

The transformations~(\ref{BRSTExSQ}) realize the extended BRST symmetry,
while the transformations (\ref{BRSTExSQ2}) express the symmetry related to
the $Sp(2)$ invariance of the quantum action. This interpretation is
explained in the following section, by the relation of the present formalism
to the $Sp(2)$ covariant superfield scheme \cite{l} and the $osp(1,2)$
covariant approach \cite{glm}. Note that the admissibility condition (\ref
{IV51}) is not required for the proof of invariance. As will be shown in the
following section, this condition establishes the relation between the
proposed formalism and the quantization schemes \cite{glm,l}.

The transformations of extended BRST symmetry (\ref{BRSTExSQ}) permit
establishing the independence of the vacuum functional (\ref{ZExSQ}) from a
choice of the gauge Boson $F(\Phi )$. Indeed, any infinitesimal change $%
F\rightarrow F+\delta F$ can be compensated by a change of variables (\ref
{BRSTExSQ}) with the parameters $\mu _{a}=-(i/2\hbar )\varepsilon
_{ab}U^{b}\delta F(\Phi )$, and therefore $Z_{F+\delta F}=Z_{F}$, which
implies the independence of the $S$-matrix from the choice of gauge within
the proposed formalism.

\section{Component Analysis}

Let us consider the component representation of the formalism proposed in
the previous section in order to establish its relation with the $osp(1,2)$
covariant approach \cite{glm} and the $Sp(2)$ covariant superfield scheme 
\cite{l}.

The component form of superfields $\Phi ^{A}(\theta )$ and supersources $%
\bar{\Phi}_{A}(\theta )$ reads 
\begin{eqnarray*}
\Phi ^{A}(\theta ) &=&\phi ^{A}+\pi ^{Aa}\theta _{a}+\lambda ^{A}\theta ^{2},
\\
\bar{\Phi}_{A}(\theta ) &=&\bar{\phi}_{A}-\theta ^{a}\phi _{Aa}^{\ast
}-\theta ^{2}\eta _{A}.
\end{eqnarray*}
Here, the components $(\phi ^{A},\pi ^{Aa},\lambda ^{A},\bar{\phi}_{A},\phi
_{Aa}^{\ast },\eta _{A})$ are identical with the set of variables used for
the construction of the vacuum functional in the quantization schemes \cite
{glm,l}.

By virtue of the manifest structure of $\Phi ^{A}(\theta )$, $\bar{\Phi}%
_{A}(\theta )$, the component representation of the integration measure in (%
\ref{ZExSQ}) is given by 
\begin{equation}
d\Phi \,d\bar{\Phi}=d\phi \,d\pi \,d\lambda \,d\bar{\phi}\,d\phi ^{\ast
}\,d\eta ,  \label{mes}
\end{equation}
and the functional $\bar{\Phi}\Phi $ in (\ref{phi_phi}) has the form 
\begin{equation}
\bar{\Phi}\Phi =\bar{\phi}_{A}\lambda ^{A}+\phi _{Aa}^{\ast }\pi ^{Aa}-\eta
_{A}\phi ^{A}.  \label{phi^2}
\end{equation}

Let us denote $F(\Phi ,\bar{\Phi})\equiv \tilde{F}(\phi ,\pi ,\lambda ,\bar{%
\phi},\phi ^{\ast },\eta )$. Then the superbrackets $(\;\,,\;)^{a}$ and $%
\{\;\,,\;\}_{\alpha }$ in (\ref{bracket}) acquire the following component
structure: 
\begin{eqnarray}
(F,G)^{a} &=&\frac{\delta \tilde{F}}{\delta \phi ^{A}}\frac{\delta \tilde{G}%
}{\delta \phi _{Aa}^{\ast }}+\varepsilon ^{ab}\frac{\delta \tilde{F}}{\delta
\pi ^{Ab}}\frac{\delta \tilde{G}}{\delta \bar{\phi}_{A}}-(\tilde{F}%
\leftrightarrow \tilde{G})\;(-1)^{(\varepsilon (F)+1)(\varepsilon (G)+1)}, 
\nonumber \\
\{F,G\}_{\alpha } &=&(\sigma _{\alpha })_{B}^{~~\!A}\left( \frac{\delta 
\tilde{F}}{\delta \phi ^{A}}\frac{\delta \tilde{G}}{\delta \eta _{B}}+\frac{%
\delta \tilde{F}}{\delta \lambda ^{A}}\frac{\delta \tilde{G}}{\delta \bar{%
\phi}_{B}}\right) +\left( \frac{\delta \tilde{F}}{\delta \pi ^{Ab}}(\sigma
_{\alpha })_{\;\;a}^{b}+\frac{\delta \tilde{F}}{\delta \pi ^{Ba}}(\sigma
_{\alpha })_{\;\;A}^{B}\right) \frac{\delta \tilde{G}}{\delta \phi
_{Aa}^{\ast }}  \nonumber \\
&&+(\tilde{F}\leftrightarrow \tilde{G})(-1)^{\varepsilon (F)\varepsilon (G)},
\label{bracket2}
\end{eqnarray}
while the second-order operators $\Delta ^{a}$ and $\Delta _{\alpha }$ in (%
\ref{DeltaaExSQ2}) take the form 
\begin{eqnarray}
\Delta ^{a} &=&(-1)^{\varepsilon _{A}}\frac{\delta _{l}}{\delta \phi ^{A}}%
\frac{\delta }{\delta \phi _{Aa}^{\ast }}+(-1)^{\varepsilon
_{A}+1}\varepsilon ^{ab}\frac{\delta _{l}}{\delta \pi ^{Ab}}\frac{\delta }{%
\delta \bar{\phi}_{A}}\,,  \nonumber \\
\Delta _{\alpha } &=&(-1)^{\varepsilon _{A}}(\sigma _{\alpha
})_{B}^{~~\!A}\left( \frac{\delta _{l}}{\delta \phi ^{A}}\frac{\delta }{%
\delta \eta _{B}}+\frac{\delta _{l}}{\delta \lambda ^{A}}\frac{\delta }{%
\delta \bar{\phi}_{B}}\right)  \nonumber \\
&&+(-1)^{\varepsilon _{A}+1}\left( \frac{\delta _{l}}{\delta \pi ^{Ab}}%
(\sigma _{\alpha })_{\;\;a}^{b}+\frac{\delta _{l}}{\delta \pi ^{Ba}}(\sigma
_{\alpha })_{\;\;A}^{B}\right) \frac{\delta }{\delta \phi _{Aa}^{\ast }}\,.
\label{delta2}
\end{eqnarray}
In (\ref{U&V}), the first-order operators $V^{a}$ and $V_{\alpha }$ have the
component representation 
\begin{eqnarray}
V^{a} &=&\varepsilon ^{ab}\phi _{Ab}^{\ast }\frac{\delta }{\delta \bar{\phi}%
_{A}}-\eta _{A}\frac{\delta }{\delta \phi _{Aa}^{\ast }}\,,  \nonumber \\
V_{\alpha } &=&\bar{\phi}_{B}(\sigma _{\alpha })_{~~\!A}^{B}\frac{\delta }{%
\delta \bar{\phi}_{A}}\,+\left( \phi _{Ab}^{\ast }(\sigma _{\alpha
})_{~a}^{b}+\phi _{Ba}^{\ast }(\sigma _{\alpha })_{~~\!A}^{B}\right) \frac{%
\delta }{\delta \phi _{Aa}^{\ast }}+\eta _{B}(\sigma _{\alpha })_{~~\!A}^{B}%
\frac{\delta }{\delta \eta _{A}}\,,  \label{Valpha}
\end{eqnarray}
while the first-order operators $U^{a}$ and $U_{\alpha }$ are given by 
\begin{eqnarray}
U^{a} &=&(-1)^{\varepsilon _{A}}\varepsilon ^{ab}\lambda ^{A}\frac{\delta
_{l}}{\delta \pi ^{Ab}}-(-1)^{\varepsilon _{A}}\pi ^{Aa}\frac{\delta _{l}}{%
\delta \phi ^{A}}\,,  \nonumber \\
U_{\alpha } &=&\phi ^{B}(\sigma _{\alpha })_{B}^{~~\!A}\frac{\delta _{l}}{%
\delta \phi ^{A}}+\left( \pi ^{Ab}(\sigma _{\alpha })_{b}^{~a}+\pi
^{Ba}(\sigma _{\alpha })_{B}^{~~\!A}\right) \frac{\delta _{l}}{\delta \pi
^{Aa}}+\lambda ^{B}(\sigma _{\alpha })_{B}^{~~\!A}\frac{\delta _{l}}{\delta
\lambda ^{A}}\,.  \label{Ualpha}
\end{eqnarray}
Finally, the component form of the admissibility condition (\ref{IV51}) 
\begin{equation}
\frac{\delta \tilde{W}}{\delta \eta _{A}}=\phi ^{A}  \label{adm3}
\end{equation}
implies a simplification of the quantum action: 
\begin{equation}
\tilde{W}={\cal W}(\phi ,\lambda ,\pi ,\bar{\phi},\phi ^{\ast })+\eta
_{A}\phi ^{A}.  \label{simpl}
\end{equation}

To establish the relation between the proposed superfield scheme and the $%
osp(1,2)$ covariant formalism \cite{glm}, we note, first of all, that the
operators $U_{\alpha }$ and $V_{\alpha }$ in (\ref{Valpha}), (\ref{Ualpha})%
{\bf \ }coincide with the generators of $Sp(2)$ invariance \cite{glm}. In
particular, equation (\ref{cond2}) is the condition of $Sp(2)$ invariance
for the gauge Boson $\tilde{F}(\phi ,\pi ,\lambda )$.

Let us subject the quantum action $\tilde{W}$ to the restrictions

\begin{equation}
\frac{\delta \tilde{W}}{\delta \lambda ^{A}}=\frac{\delta \tilde{W}}{\delta
\pi ^{Aa}}=0,  \label{w1}
\end{equation}
reducing the variables of $\tilde{W}$ to the set $(\phi ^{A},\bar{\phi}%
_{A},\phi _{Aa}^{\ast },\eta _{A})$, parameterizing the quantum action in
the $osp(1,2)$ covariant scheme. By virtue of (\ref{w1}) and the component
representations (\ref{bracket2})--(\ref{Ualpha}), the set of equations (\ref
{GEq1ExSQ}), (\ref{GEq1ExSQ2}) becomes identical to the massless limit of
the master equations in the $osp(1,2)$ covariant formalism.

Using the conditions (\ref{adm3}), (\ref{w1}), with allowance for the
properties of the matrices $\sigma _{\alpha }$ in (\ref{sigma}), (\ref
{propsigma}), one can transform the subsidiary master equations (\ref
{GEq1ExSQ2}) into the condition of $Sp(2)$ invariance for the quantum action 
\cite{glm} 
\[
(\sigma _{\alpha })_{B}^{~~\!A}\frac{\delta \tilde{W}}{\delta \phi ^{A}}\phi
^{B}+V_{\alpha }\tilde{W}=0, 
\]
which thus establishes the interpretation of the symmetry transformations (%
\ref{BRSTExSQ2}) related to equations (\ref{GEq1ExSQ2}).

Let us also restrict the gauge-fixing Boson to the class of gauges used in
the $osp(1,2)$ covariant scheme: $\tilde{F}=\tilde{F}(\phi )$. Then, with
allowance for the component form (\ref{Ualpha}) of the operators $U_{\alpha
} $, the condition of $Sp(2)$ invariance (\ref{cond2}) reduces to 
\begin{equation}
(\sigma _{\alpha })_{B}^{~~\!A}\frac{\delta \tilde{F}}{\delta \phi ^{A}}\phi
^{B}=0,  \label{sp2}
\end{equation}
which, in view of the admissibility condition (\ref{adm3}), can be rewritten
as 
\begin{equation}
(\sigma _{\alpha })_{B}^{~~\!A}\frac{\delta \tilde{F}}{\delta \phi ^{A}}%
\frac{\delta \tilde{W}}{\delta \eta _{B}}=0.  \label{equiv}
\end{equation}
Equations (\ref{sp2}) and (\ref{equiv}) reproduce the whole set of
subsidiary conditions used in the $osp(1,2)$ covariant scheme to provide an $%
Sp(2)$ invariant gauge-fixing \cite{glm}.

Let us establish the relation of the vacuum functional (\ref{ZExSQ}), given
in terms of $\tilde{W}=\tilde{W}(\phi ,\bar{\phi},\phi ^{\ast },\eta )$ and $%
\tilde{F}=\tilde{F}(\phi )$, to the vacuum functional of the $osp(1,2)$
covariant scheme \cite{glm}. Using the component form (\ref{Ualpha}) of the
operators $U^{a}$, and integrating out the variables $\eta _{A}$, with
allowance for (\ref{mes}), (\ref{phi^2}), (\ref{simpl}), we can represent
the vacuum functional (\ref{ZExSQ}) in the form 
\begin{equation}
Z=\int d\phi \,d\phi ^{\ast }\,d\pi \,d\bar{\phi}\,d\lambda \,\exp \left[ 
\frac{i}{\hbar }\left( {\cal W}+{\cal X}+\bar{\phi}_{A}\lambda ^{A}+\phi
_{Aa}^{\ast }\pi ^{Aa}\right) \right] ,  \label{Z1}
\end{equation}
where the functional $\tilde{W}={\cal W}+\eta _{A}\phi ^{A}$ satisfies (\ref
{GEq1ExSQ}), (\ref{GEq1ExSQ2}), (\ref{adm3}), and the gauge-fixing term $%
{\cal X}$ is given by 
\[
{\cal X}=-\frac{\delta \tilde{F}}{\delta \phi ^{A}}\lambda ^{A}-\frac{1}{2}%
\varepsilon _{ab}\pi ^{Aa}\frac{\delta ^{2}\tilde{F}}{\delta \phi ^{A}\delta
\phi ^{B}}\pi ^{Bb}, 
\]
with $\tilde{F}$ subject to (\ref{sp2}). On the other hand, the vacuum
functional in the massless limit of the $osp(1,2)$ covariant formalism \cite
{glm} can be represented as 
\begin{equation}
Z=\int d\phi \,{\exp }\left( \frac{i}{\hbar }S_{\,{\rm eff}}\right) ,
\label{ospvac}
\end{equation}
\[
\left. S_{\,{\rm eff}}(\phi )=S_{\,{\rm ext}}(\phi ,\bar{\phi},\phi ^{\ast
},\eta )\right| _{\bar{\phi}=\phi ^{\ast }=\eta =0},\;\;\;{\rm exp}\left[
(i/\hbar )S_{\,{\rm ext}}\right] =\hat{U}(Y)\,{\rm exp}\left[ (i/\hbar )S%
\right] . 
\]
Here, $S=S(\phi ,\bar{\phi},\phi ^{\ast },\eta )$ is the quantum action
subject to the system of master equations and subsidiary conditions (\ref
{GEq1ExSQ}), (\ref{GEq1ExSQ2}), (\ref{adm3}) satisfied by $\tilde{W}=\tilde{W%
}(\phi ,\bar{\phi},\phi ^{\ast },\eta )$, while $\hat{U}(Y)$ is an operator
of the form 
\[
\hat{U}(Y)={\rm exp}\left( \frac{\delta Y}{\delta \phi ^{A}}\frac{\delta }{%
\delta \bar{\phi}_{A}}+\frac{i\hbar }{2}\varepsilon _{ab}\frac{\delta }{%
\delta \phi _{Aa}^{\ast }}\frac{\delta ^{2}Y}{\delta \phi ^{A}\delta \phi
^{B}}\frac{\delta }{\delta \phi _{Bb}^{\ast }}\right) , 
\]
where $Y=Y(\phi )$ is a gauge-fixing Boson restricted by the same condition
of $Sp(2)$ invariance (\ref{sp2}) which is imposed on $\tilde{F}=\tilde{F}%
(\phi )$. To establish the identity between the vacuum functionals (\ref{Z1}%
) and (\ref{ospvac}), it is sufficient to set $S=\tilde{W}$ and $Y=\tilde{F}$%
.

Let us finally establish the relation of the proposed superfield scheme to
the original $Sp(2)$ covariant superfield formalism \cite{l}. First, note
that the operators $U^{a}$, $V^{a}$ (\ref{U&V}), which also appear in the
symmetry transformations (\ref{BRSTExSQ}), are naturally interpreted \cite{l}
as generators of transformations induced by supertranslations, $\theta
^{a}\rightarrow \theta ^{a}+\mu ^{a}$. Next, the form of $(\;,\;)^{a}$ and $%
\Delta ^{a}$ in (\ref{bracket}), (\ref{DeltaaExSQ2}) implies that equations (%
\ref{GEq1ExSQ}) are identical with the master equations of the approach \cite
{l}. Then, the admissibility condition (\ref{adm3}) and the related
dependence (\ref{simpl}) of $\tilde{W}$ on the variables $\eta ^{A}$, with
allowance for (\ref{mes}){\bf , }(\ref{phi^2}), permit us to rewrite the
vacuum functional (\ref{ZExSQ}) in the form

\begin{equation}
Z=\int d\Phi \,d\bar{\Phi}\,\rho (\bar{\Phi})\exp \left[ \frac{i}{\hbar }%
\left( W(\Phi ,\bar{\Phi})-\frac{1}{2}\varepsilon _{ab}U^{a}U^{b}F(\Phi )+%
\bar{\Phi}\Phi \right) \right] ,  \label{vacs(2)}
\end{equation}
where $\rho (\bar{\Phi})$ is an integration weight, given by 
\[
\rho (\bar{\Phi})=\delta \left( \int d^{2}\theta \,\bar{\Phi}(\theta
)\right) =\delta \left( \eta \right) . 
\]
The integral (\ref{vacs(2)}) is identical with the vacuum functional of the $%
Sp(2)$\ covariant superfield scheme \cite{l}, where the corresponding
objects $W(\Phi ,\bar{\Phi})$ and $F(\Phi )$ are subject to additional
restrictions, (\ref{GEq1ExSQ2}), (\ref{IV51}), (\ref{cond2}), which ensure
the $Sp(2)$ invariance of the quantum theory. Note that the symmetry
transformations (\ref{BRSTExSQ}) related to the master equations (\ref
{GEq1ExSQ}) coincide with the superfield form of extended BRST\ symmetry 
\cite{l} in terms of supertranslations.

\section{Summary}

The present work is motivated by the problem of a consistent superspace
formulation of extended BRST\ symmetry on the basis of the $osp(1,2)$
covariant quantization scheme \cite{glm} for general gauge theories. Here,
by a consistent superspace formulation we understand a superfield
quantization scheme in which extended BRST\ symmetry, realized in terms of
supertranslations, is completely controlled by the quantum master equations
(see, e.g., \cite{l}). This consistency condition requires \cite{lm2} that a
superfield $osp(1,2)$ covariant scheme should contain the $Sp(2)$ covariant
superfield formalism \cite{l} in the limit of a vanishing mass (a parameter
introduced to provide an $Sp(2)$ invariant renormalization \cite{glm}),
which arises in the superalgebra \cite{glm} of generating operators of
quantum master equations. The fulfillment of the above requirement turns out
to be a non-trivial problem (see, e.g., \cite{lm2}), related to a
realization of the $osp(1,2)$ superalgebra of generating operators in a form
compatible with the superfield antibrackets used in \cite{l}. To approach
this problem, we propose a superfield scheme which can be regarded as the
massless limit of a consistent superspace formulation of the $osp(1,2)$
covariant formalism. Namely, we propose a modification of the $Sp(2)$
covariant superfield scheme \cite{l} on the basis of a superalgebra of
generating operators isomorphic to the massless limit of the corresponding
superalgebra of $osp(1,2)$ covariant quantization \cite{glm}. The
realization of generating operators is consistent with the superfield
antibrackets \cite{l}. As a result, the superspace form of extended BRST\
symmetry is encoded by the quantum master equations without imposing
restrictions eliminating superfield components (cf. \cite{lm2}). An
additional admissibility condition reduces the formalism to the original $%
Sp(2)$ covariant superfield scheme and to the massless limit of the $%
osp(1,2) $ covariant scheme in particular cases of gauge-fixing and
solutions of the master equations. Analysis of massive extensions of the
proposed scheme (as well as the study of its possible arbitrariness) may
provide a constructive way of finding a consistent superspace formulation of
the $osp(1,2)$ covariant approach. It appears interesting to extend the
consideration of the present work to the superfield scheme \cite{gm}, where
a superspace description of $osp(1,2)$ covariant quantization is proposed by
considering the $osp(1,2)$ superalgebra as a subalgebra of the $sl(1,2)$
superalgebra \cite{alg}, which can be regarded as the algebra of conformal
generators in a superspace with two anticommuting coordinates. The approach 
\cite{gm} suggests an intriguing possibility to realize extended BRST
symmetry in terms of conformal transformations in superspace, whereby
supertranslations \cite{l} are included as a particular case.

{\bf Acknowledgement:}{\large \ }D.M.G. is grateful to the foundations
FAPESP, CNPq and DAAD for support. The work of P.Yu.M. was supported by
FAPESP, grant 02/00423-4.


\begin{thebibliography}{99}
\bibitem{blt}  I.A.~Batalin, P.M.~Lavrov and I.V.~Tyutin, {\it J. Math.
Phys.\/} {\bf 31}, 1487 (1990); {\bf 32}, 532 (1991); {\bf 32}, 2513 (1991).

\bibitem{3pl}  I.A.~Batalin and R.~Marnelius, {\it Phys. Lett.} {\bf B350}
(1995) 44; {\it Nucl. Phys.} {\bf B465} (1996) 521;\newline
I.A.~Batalin, R.~Marnelius and A.M.~Semikhatov, {\it Nucl. Phys.} {\bf B446}
(1995) 249.\newline
B.~Geyer, D.M.~Gitman and P.M.~Lavrov, {\it Mod. Phys. Lett.} {\bf A14}
(1999) 661.

\bibitem{glm}  B.~Geyer, P.M.~Lavrov and D.~M\"{u}lsch, {\it Acta Phys.
Polon.} {\bf B29}, 2637 (1998); {\it J.~Math.~Phys.} {\bf 40}, 674 (1999); 
{\bf 40}, 6189 (1999).

\bibitem{brst}  C.~Becchi, A.~Rouet and R.~Stora, {\it Phys. Lett.\/} {\bf %
B52}, 344 (1974);\newline
I.V.~Tyutin, {\it Gauge Invariance in Field Theory and Statistical Mechanics}%
, Lebedev Inst. preprint No. 39 (1975).

\bibitem{antibrst}  G.~Curci and R.~Ferrari, {\it Phys. Lett.\/} {\bf B63},
91 (1976);

I.~Ojima, {\it Prog. Theor. Phys. Suppl.\/} {\bf 64}, 625 (1979).

\bibitem{bpt}  L.~Bonora and M.~Tonin, {\it Phys. Lett. }\/{\bf B98}, 48
(1981);\newline
L.~Bonora, P.~Pasti and M.~Tonin, {\it J. Math. Phys.}\/ {\bf 23}, 839
(1982).

\bibitem{l}  P.M.~Lavrov, {\it Phys. Lett. \/}{\bf B366}, 160 (1996); {\it %
Theor. Math. Phys. \/}{\bf 107}, 602 (1996);\newline
P.M.~Lavrov and P.Yu.~Moshin, {\it Phys. Lett.} {\bf B508}, 127 (2001); {\it %
Theor. Math. Phys. }{\bf 129 }1645 (2001).

\bibitem{gm}  B.~Geyer and D.~M\"{u}lsch, {\it J. Math. Phys.\/} {\bf 41},
7304 (2000).

\bibitem{lm2}  P.M.~Lavrov and P.Yu.~Moshin, {\it Grav. Cosmol.} {\bf 8}, 49
(2002).

\bibitem{alg}  A.~Pais and V.~Rittenberg, {\it J. Math. Phys.\/} {\bf 16},
2062 (1975);\newline
M.~Scheunert, W.~Nahm and V.~Rittenberg, {\it J. Math. Phys.\/} {\bf 18},
146 (1976);\newline
F.A.~Berezin and V.N.~Tolstoy, {\it Commun. Math. Phys.} {\bf 78}, 409
(1981);\newline
L.~Frappat, P.~Sorba and A.~Sciarrino, {\it Dictionary on Lie superalgebras,}
ENSLAPP-AL-600/96; hep-th/9607161.
\end{thebibliography}
\end{document}